\documentclass{elsart}
\usepackage{amssymb}
\usepackage{amsmath}
\usepackage[dvips]{graphicx}

\journal{: \  International Journal of
Engineering Science  \textbf{47},  n$^o$ 5-6, pp. 691-699 (2009)\qquad\qquad\qquad\qquad\qquad\qquad\qquad\qquad\qquad\qquad}
\input{tcilatex}
\begin{document}

\begin{frontmatter}

\title{\large Liquid Nanofilms.  \\
 A Mechanical Model for the Disjoining Pressure}

\author{Henri Gouin}
\ead{henri.gouin@univ-cezanne.fr}

\address{University of Aix-Marseille \& M2P2, C.N.R.S.  U.M.R.  6181, \\ Case 322, Av. Escadrille
Normandie-Niemen, 13397 Marseille Cedex 20 France}

\begin{abstract}
Liquids in contact with solids are submitted to intermolecular
forces  making liquids   heterogeneous and,  in a mechanical model,
the stress tensor is not any more spherical as in homogeneous bulks.
  The aim  of this article is to show that  a square-gradient functional
  taking into account  the volume
liquid  free energy  corrected with two surface liquid density
functionals is a mean field approximation allowing to study
structures of very thin liquid nanofilms near plane solid walls. The
model determines analytically the concept of disjoining pressure for
liquid films of  thicknesses  of a very few number of nanometers and
yields a behavior in good agreement with the shapes of  experimental
curves carried out by Derjaguin and his successors.

\end{abstract}

\vskip 0.5cm

\end{frontmatter}
\vskip 2.4cm

\section{Introduction}

The technical development of sciences allows us to observe phenomena at
length scales of a very few number of nanometers. This \emph{nanomechanics}
infers applications in numerous fields, including medicine and biology. It
reveals new behaviors, often surprising and essentially different from those
that are usually observed at macroscopic and also at microscopic scales \cite%
{Bhushan}. Currently simple models proposing realistic qualitative behaviors
need to be developed in different fields of nanosciences even if their
comparison with experimental data may be criticized at a quantitative level.%
\newline
As pointed out in experiments with water, the density of liquid water is
found to be changed in narrow pores \cite{Karasev}. The first reliable
evidence of this effect was reported by V.V. Karasev, B.V. Derjaguin and
E.N. Efremova in 1962 and found after by many others (\cite{Derjaguin},
pages 240-244). In order to evaluate the structure of thin interlayers of
water and other liquids, Green-Kelly and Derjaguin employed a method based
on measuring changes in birefringence \cite{Green-Kelly}; they found
significant anisotropy of water interlayers. \newline
In a recent article, the equations of motion of thin films were considered
by taking into account the variation of the disjoining pressure along the
layer \cite{gouinijes}. The aim of this paper is to study, by means of a
continuous mechanical model, the disjoining pressure and the behavior for
very thin liquid films at the \emph{mesoscopic scale} of a few number of
nanometers.

Since van der Waals at the end of the nineteenth century, the fluid
inhomogeneities in liquid-vapor interfaces were represented in continuous
models by taking into account a volume energy depending on space density
derivative \cite{Slemrod,Seppecher,Widom,Kaz,Onuki}. Nevertheless, the
corresponding square-gradient functional is unable to model repulsive force
contributions and misses the dominant damped oscillatory packing structure
of liquid interlayers near a substrate wall \cite{chernov1,Weiss}.
Furthermore, the decay lengths are correct only close to the liquid-vapor
critical point where the damped oscillatory structure is subdominant \cite%
{Evans1}. In mean field theory, weighted density-functional has been used to
explicitly demonstrate the dominance of this structural contribution in van
der Waals thin films and to take into account long-wavelength capillary-wave
fluctuations as in papers that renormalize the square-gradient functional to
include capillary wave fluctuations \cite{Fisher}. In contrast, fluctuations
strongly damp oscillatory structure and it is mainly for this reason that
van der Waals' original prediction of a \emph{hyperbolic tangent} is so
close to simulations and experiments \cite{rowlinson}.\newline
To propose an analytic expression in density-functional theory for liquid
film of a very few nanometer thickness near a solid wall, we add a liquid
density-functional at the solid surface and a surface density functional at
the liquid-vapor interface to the square-gradient functional representing
the volume free energy of the fluid. This kind of functional is well-known
in the literature \cite{Fisher1}. It was used by Cahn in a phenomenological
form, in a well-known paper studying wetting near a critical point \cite%
{Cahn0}. An asymptotic expression is obtained in \cite{gouin} with an
approximation of hard sphere molecules and London potentials for
liquid-liquid and solid-liquid interactions: in this way, we took into
account the power-law behavior which is dominant in a thin liquid film in
contact with a solid.\newline
The disjoining pressure $\Pi$ is a well adapted tool for a very thin liquid
film of thickness $h$. In cases of Lifshitz analysis \cite{Lifshitz} and van
der Waals theory, the disjoining pressure behaviors are respectively as $%
\Pi\sim h^{-3}$ and $\Pi\sim \exp ( -h)$. None of them represents correctly
experimental results for a film with a thickness ranging over a few
nanometers.\newline
Then, the gradient expansion missing the physically dominant damped
oscillatory packing structure of the liquid near a substrate wall and only
working up a smooth exponential decay is corrected by surface energies
issued from London forces which model power-law dispersion interaction:
since the only structure that the square gradient functional can yield is
monotonic exponential, the surface energies take account of the eventual
dominance of attractive power-law dispersion interactions. In fact power-law
wings are physically present for liquid film of several nanometers and it is
the reason we propose a study only for films in a range of few nanometers.

\section{The density-functional}

In our model, the free energy density-functional of an inhomogeneous liquid
in a domain $O$ of boundary $\partial O$ is taken in the general form
\begin{equation}
F = \int\int\int_O \varepsilon\ dv + \int\int_{\partial O} \varphi\ ds ,
\label{density functional}
\end{equation}
where $\varepsilon$ is the specific free energy and $\varphi$ is a generic
surface free energy of $\partial O$.\newline
In our problem, we consider a horizontal plane liquid layer $(L)$ contiguous
to its vapor bulk and in contact with a plane solid wall $(S)$; the z-axis
is perpendicular to the solid surface $(S)$. The liquid film thickness is
denoted by $h$. Far from its critical point, the liquid at level $z=h$ is
situated at a distance order of two molecular diameters from the vapor bulk
and the liquid-vapor interface is assimilated to a surface $(\Sigma)$ at $%
z=h $. Then, the free energy density-functional (\ref{density functional})
gets the particular form
\begin{equation}
F = \int\int\int_{(L)} \varepsilon\ dv + \int\int_{(S)} \phi\ ds +
\int\int_{(\Sigma)} \psi\ ds.  \label{density functional2}
\end{equation}
where $\varphi$ is shared in two parts $\phi$ and $\psi$ respectively
associated with $(S)$ and $(\Sigma)$.

$\bullet\quad $ In Rel. (\ref{density functional2}), the \emph{first integral%
} (energy of volume $(L)$) is associated with square-gradient approximation
when we introduce a specific free energy of the fluid
\begin{equation*}
\varepsilon =\varepsilon(\rho,\beta)  \label{specific energy}
\end{equation*}
at a given temperature $\theta$ as a function of density $\rho$ and $\beta= (%
\mathrm{grad\, \rho)^2}$. Specific free energy $\varepsilon$ characterizes
both fluid properties of \emph{compressibility} and \emph{molecular
capillarity} of liquid-vapor interfaces. In accordance with gas kinetic
theory \cite{Rocard}, scalar $\lambda =2\rho\, \varepsilon _{,\beta }(\rho,
\beta)$ (where $\varepsilon _{,\beta }$ denotes the partial derivative with
respect to $\beta$) is assumed to be constant at a given temperature and
\begin{equation}
\rho \,\varepsilon =\rho \,\alpha (\rho)+\frac{\lambda }{2}\,(\text{grad\ }%
\rho )^{2},  \label{internal energy}
\end{equation}
where the term $({\lambda }/{2})\,(\mathrm{grad\ \rho )^{2}}$ is added to
the volume free energy $\rho \,\alpha (\rho)$ of a compressible homogeneous
fluid. We denote the pressure term associated with specific free energy $%
\alpha(\rho)$ by
\begin{equation}
P(\rho)=\rho ^{2}\alpha^{\,\prime }(\rho ) .  \label{pressure}
\end{equation}
$\bullet\quad $ In Rel. (\ref{density functional2}), the \emph{second
integral} (energy of surface $(\Sigma)$) is defined through a model of
molecular interactions between the fluid and the solid wall. In fact, near a
solid wall, the London potentials of liquid-liquid and liquid-solid
interactions are
\begin{equation*}
\left\{
\begin{array}{c}
\displaystyle\;\;\;\;\;\;\varphi _{ll}=-\frac{c_{ll}}{r^{6}}\;,\text{ \
when\ }r>\sigma _{l}\;\;\text{and }\;\ \varphi _{ll}=\infty \text{ \ when \ }%
r\leq \sigma _{l}\, ,\  \\
\displaystyle\;\;\;\;\;\;\varphi _{ls}=-\frac{c_{ls}}{r^{6}}\;,\text{ \
when\ }r>\delta \;\;\text{and }\;\ \varphi _{ls}=\infty \text{ \ when \ }%
r\leq \delta \;, \
\end{array}
\right.
\end{equation*}
where $c_{ll}$ and $c_{ls}$ are two positive constants, $\sigma _{l}$ and $%
\sigma _{s}$ denote liquid (fluid) and solid molecular diameters, $\delta =%
\frac{1}{2}(\sigma _{l}+\sigma _{s})$ is the minimal distance between
centers of liquid and solid molecules \cite{Israel}. In the theory of
additive and non-retarded molecular interactions, coefficients $c_{ll}$ and $%
c_{ls}$ are connected with Hamaker constants $A_{ll}$ and $A_{ls} $ through
the relations $A_{ll} = \pi^2 c_{ll}\rho_l^2$ and $A_{ls} = \pi^2
c_{ls}\rho_l\rho_{s}$, where $\rho_l$ and $\rho_{s}$ respectively denote
liquid bulk and solid densities \cite{Hamak}. Forces between liquid and
solid have short range and can be simply described by adding a special
energy at the surface. This is not the entire interfacial energy: another
contribution comes from the distortions in the liquid density profile near
the wall \cite{gouin,de Gennes2}. Finally, for a plane solid wall (at a
molecular scale), this surface free energy is obtained in the form
\begin{equation}
\phi(\rho)=-\gamma _{1}\rho+\frac{1}{2}\,\gamma_{2}\,\rho^{2}.
\label{surface energy}
\end{equation}
Here $\rho$ denotes the liquid density value at the wall. The constants $%
\gamma _{1}$, $\gamma _{2}$ are positive and given by the relations $%
\displaystyle
\gamma _{1}=\frac{\pi c_{ls}}{12\delta ^{2}m_{l}m_{s}} \rho_{s}$, $%
\displaystyle
\gamma _{2}=\frac{\pi c_{ll}}{12\delta^2 m_{l}^{2}}$\,, where $m_{l}$ et $%
m_{s}$ respectively denote the masses of liquid (fluid) and solid molecules
\cite{gouin}. Moreover, we have $\displaystyle
\lambda = \frac{2\pi c_{ll}}{3\sigma_l \,m_{l}^{2}}$\,.\newline
$\bullet\quad $ In Rel. (\ref{density functional2}), let us consider the
\emph{third integral}. The conditions in the vapor bulk are $\displaystyle
\mathrm{grad}\, \rho =0$ and $\Delta \rho = 0$ with $\Delta$ denoting the
Laplace operator. Far from the critical point, a way to compute the total
free energy of the complete liquid-vapor layer is to add the energy of the
liquid layer $(L)$ located between $z=0$ and $z=h$ (first integral of Rel. (%
\ref{density functional2})), the surface energy of the solid wall $(S)$ at $%
z=0$ (second integral of Rel. (\ref{density functional2})), the energy of
the liquid-vapor interface of a few Angstr\"{o}m thickness assimilated to a
surface $(\Sigma)$ at $z=h$ and the energy of the vapor layer located
between $z=h$ and $z=+\infty$ \cite{gavrilyuk}. The liquid at level $z=h$ is
situated at a distance order of two molecular diameters from the vapor bulk
and the vapor has a negligible density with respect to the liquid density
\cite{pismen}. In our model, these two last energies can be expressed by
writing a unique energy $\psi$ per unit surface located on the mathematical
surface $(\Sigma)$ at $z=h $ : by a calculation like in \cite{gouin}, we can
write $\psi$ in a form analogous to expression (\ref{surface energy}) and
also expressed in \cite{de Gennes2} in the form $\psi(\rho) = -\gamma
_{5}\rho+\frac{1}{2}\,\gamma_{4}\,\rho^{2}$; but with a \emph{wall}
corresponding to a \emph{negligible density}, $\gamma _{5}\simeq 0$ and the
surface free energy $\psi$ is reduced to
\begin{equation}
\psi (\rho )=\frac{\gamma _{4}}{2}\ \rho ^{2} ,  \label{cl2}
\end{equation}
where $\rho$ is the liquid density at level $z=h$ and $\gamma_4$ is
associated with a distance of the order of the fluid molecular diameter
(when $\sigma_l \simeq \delta$, then $\gamma_4\simeq \gamma_2 $).
Consequently, due to the small vapor density, the surface free energy $\psi$
is the same as the one of a liquid in contact with a vacuum and expressed by
the third integral of Rel. (\ref{density functional2}).

Such a form of density functional restricted to the first two integrals was
primary expressed by Cahn; Cahn's study used a graphic representation where
energy integrals were presented as different areas in an energy-density
plane \cite{Cahn0}. Analytical computations were also tested in \cite%
{gavrilyuk} but without taking account of a complete volume free energy in
form (\ref{internal energy}).\newline
With our previous functional approximation, we obtain the equations of
equilibrium (or motion) and boundary conditions for a thin liquid film
damping a solid wall. We can compute the liquid layer thickness. The normal
stress vector acting on the wall remains constant through the layer and
corresponds to the gas-vapor bulk pressure which is usually the atmospheric
pressure.\newline
We obtain analytical results expressing the profile of density of very thin
layer at a mesoscopic scale. We deduce an analytic expression of the
disjoining pressure computed for different solid materials in contact with
nanometer scale liquid layers. For all I know, such results have not been
obtained in the literature by using both a continuous mechanical model and a
differential equation system. \newline
It is wondering to observe that the density-functional theory expressed by a
simple model correcting van der Waals' one with surface density-functionals
at the wall and the interface, enables to obtain a representation of the
disjoining pressure for very thin films which fits in with experiments by
Derjaguin and others. This result is obtained without too complex weighted
density-functionals and without taking account of quantum effects
corresponding to an Angstr\"{o}m length scale. So, this kind of functional
may be a good tool to analytically study liquids in contact with solids at a
very small nanoscale range.

\section{Equation of motion and boundary conditions}

In case of equilibrium, functional $F$ is minimal with respect to the vector
fields of \textit{virtual displacement} classically defined (as in \cite%
{Serrin}) and yields the \emph{equation of equilibrium} of the inhomogeneous
liquid and the \emph{boundary conditions} between liquid, vapor and solid
wall. In case of motions we simply add the inertial forces and the
dissipative stresses in the equation of equilibrium (to refer to the
well-known explicit calculations, see for example \cite%
{gouinijes,gouin4,Gouin1}).

\subsection{Equation of motion}

The equation of motion is
\begin{equation}
\rho \ \mathbf{\Gamma }= \text{div}\left(\mathbf{\sigma}+\mathbf{\sigma}%
_{v}\right) -\rho\; \text{grad }\Omega ,  \label{motion0}
\end{equation}
where $\mathbf{\Gamma}$ is the acceleration vector, $\Omega $ the body force
potential, $\mathbf{\sigma }$ the stress tensor generalization and $\mathbf{%
\ \sigma }_{v}$ the viscous stress tensor,
\begin{equation*}
\mathbf{\ \sigma =}-p\,\mathbf{1}-\lambda \;\text{grad\ }\rho \ \otimes \
\text{grad }\rho ,  \label{contrainte}
\end{equation*}
where $p=\rho ^{2}\varepsilon _{,\rho }-\rho \text{ div\textrm{\ }}(\lambda
\text{ grad }\rho )$ is different from the pressure term $P$ defined in (\ref%
{pressure}). \newline
For a horizontal layer, in an orthogonal system of coordinates such that the
third coordinate $z$ is the vertical direction, all physical quantities in
the layer depend only on $z$ and the stress tensor $\mathbf{\sigma }$ of the
thin film gets the form
\begin{equation*}
\mathbf{\sigma }=\left[
\begin{array}{ccc}
a_{1} & 0 & 0 \\
0 & a_{2} & 0 \\
0 & 0 & a_{3}%
\end{array}%
\right] ,\quad \mathrm{with}\quad \left\{
\begin{array}{lll}
a_{1} & = & a_{2}=-p,\quad \displaystyle p=P(\rho )-\frac{\lambda }{2}\left(
\frac{d\rho }{dz}\right) ^{2}-\lambda\, \rho\, \frac{d^2\rho}{dz^2}\,, \\
a_{3} & = & \displaystyle{-p-\lambda \left( \frac{d\rho }{dz}\right) ^{2}.}%
\end{array}%
\right.
\end{equation*}%
Let us consider a thin film of liquid at equilibrium (gravity forces are
neglected). The equation of equilibrium is
\begin{equation}
\text{div }\mathbf{\sigma =0} \,.  \label{equilibrium1a}
\end{equation}
Equation (\ref{equilibrium1a}) yields a constant value for the eigenvalue $%
a_3$,
\begin{equation*}
p+ {\lambda } \left(\frac{d\rho }{dz}\right)^2=P_{v_b} \text{,}
\label{equilibrium1b}
\end{equation*}
or
\begin{equation*}
P+\,\frac{\lambda }{2}\left(\frac{d\rho }{dz}\right)^2-\,\lambda \rho\,
\frac{d^2\rho}{dz^2}=P_{v_b}\text{,}  \label{equilibrium1b}
\end{equation*}
where $P_{v_{b}}$ denotes the pressure $P(\rho_{v_{b}})$ in the vapor bulk,
where $\rho_{v_{b}}$ is the density of the vapor \emph{mother} bulk bounding
the liquid layer. Eigenvalues $a_1, a_2$ are not constant and depend on the
distance $z$ to the solid wall \cite{Derjaguin}. At equilibrium, general Eq.
(\ref{motion0}) is equivalent to
\begin{equation}
\text{grad}\left( \, \mu \left(\rho\right) -\lambda \Delta \rho\, \right) =0%
\mathrm{,}  \label{equilibrium2a}
\end{equation}
where $\mu $ is the chemical potential at temperature $\theta $ defined to
an unknown additive constant \cite{gouinijes,gouin4}. The chemical potential
is a function of $P$ (and $\theta$) but can be also expressed as a function
of $\rho$ (and $\theta $). At temperature $\theta$, we choose as \emph{%
reference chemical potential} $\mu _{o}=\mu _{o}(\rho)$, null for the bulks
of densities $\rho _{l}$ and $\rho _{v}$ of phase equilibrium. Due to
Maxwell rule, the volume free energy associated with $\mu _{o}$ is $%
g_{o}(\rho)-P_{o}$ where $P_{o}=P(\rho _{l})=$ $P(\rho _{v})$ is the bulk
pressure and $g_{o}(\rho)= \int_{\rho _{v}}^{\rho }\mu _{o}(\rho)\,d\rho\ $
is null for the liquid and vapor bulks of phase equilibrium. The pressure $P$
is
\begin{equation}
P(\rho)=\rho \, \mu _{o}(\rho)-g_{o}(\rho)\ +P_{o} .  \label{therm.pressure}
\end{equation}
Thanks to Eq. (\ref{equilibrium2a}), we obtain in all the fluid \emph{and
not only in the liquid layer, }
\begin{equation*}
\mu _{o}(\rho)-\lambda \Delta \rho =\mu _{{o}}(\rho _{b}) ,
\label{equilibrium2b}
\end{equation*}
where $\mu _{{o}}(\rho _{b})$ is the chemical potential value of a liquid
\emph{mother} bulk of density $\rho _{b}$ such that $\mu _{{o}}(\rho _{b})=
\mu _{{o}}(\rho_{v_{b}})$. We must emphasis that $P(\rho _{b})$ and $%
P(\rho_{v_{b}})$ are unequal as for drop or bubble bulk pressures. The
density $\rho_b$ is not a fluid density in the layer but the density in the
liquid bulk from which the layer can extend (this is why Derjaguin used the
term \emph{mother liquid} \cite{Derjaguin}, page 32).\newline
In the liquid layer $(L)$,
\begin{equation}
\lambda\,\frac{d^2\rho}{dz^2} = \mu_{b}(\rho), \quad \mathrm{with}\quad
\mu_{b}(\rho) = \mu_o(\rho)-\mu_o(\rho _{b}).  \label{equilibrium2d}
\end{equation}

\subsection{Boundary conditions}

Condition at the solid wall $(S)$ associated with the free surface energy (%
\ref{surface energy}) yields \cite{Gouin1}
\begin{equation}
\lambda \left(\frac{d\rho }{dn}\right)_{|_S}+\phi ^{\prime }(\rho)_{|_S}\ =0,
\label{cl1}
\end{equation}
where $n$ is the external normal direction to the fluid. Equation (\ref{cl1}%
) yields
\begin{equation}
\lambda \left(\frac{d\rho }{dz}\right)_{|_{z=0}}=-\gamma _{1}+\gamma _{2\
}\rho_{|_{z=0}} .  \label{BC1}
\end{equation}
The sign of $-\gamma _{1}+\gamma _{2\ }\rho_{|_{z=0}}$ determines the
wettability of the fluid on the wall: the fluid damps the solid wall when $%
\gamma _{1}-\gamma _{2}\rho_{|_{z=0}} >0$ and does not damp the solid wall
when $\gamma _{1}-\gamma _{2}\rho_{|_{z=0}} <0$ \cite{Gouin1,gouin6}.

Condition at the liquid-vapor interface $(\Sigma)$ associated with the free
surface energy (\ref{cl2}) yields
\begin{equation}
\lambda \left(\frac{d\rho }{dz}\right)_{|_{z=h}}=-\gamma _{4}\
\rho_{|_{z=h}}\,.  \label{BC2}
\end{equation}
Equation (\ref{BC2}) defines a film thickness by introducing a \emph{%
reference} point inside the liquid-vapor interface bordering the liquid
layer with a convenient density at $z=h$ \cite{gavrilyuk}.\newline
We notice that to study the stress tensor in the layer, we must also add to
conditions (\ref{BC1},\ref{BC2}) on density, the classical surface
conditions on the stress vector associated with the total stress tensor $%
\mathbf{\sigma+\sigma}_v$ \cite{Gouin1}.

\section{The disjoining pressure for horizontal liquid films}

We consider fluids and solids \emph{at a given temperature} $\theta $. The
hydrostatic pressure in a thin liquid layer located between a solid wall and
a vapor bulk differs from the pressure in the contiguous liquid phase. At
equilibrium, the additional pressure in the layer is called the \emph{%
disjoining pressure} \cite{Derjaguin}.
\begin{figure}[h]
\begin{center}
\includegraphics[width=9cm]{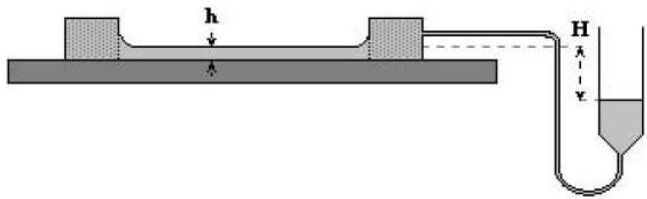}
\end{center}
\caption{\emph{Diagram of the technique for determining the disjoining
pressure isotherms of wetting films on a solid substrate: a circular wetting
film is formed on a flat substrate to which a microporous filter with a
cylindrical hole is clamped. A pipe connects the filter filled with the
liquid to a reservoir containing the liquid mother bulk that can be moved by
a micrometric device. The disjoining pressure is equal to $\Pi =(\protect%
\rho _{b}-\protect\rho _{{v_{b}}})\,gH$ (From Ref. \protect\cite{Derjaguin},
page 332).}}
\label{fig1}
\end{figure}
\begin{figure}[h]
\begin{center}
\includegraphics[width=6cm]{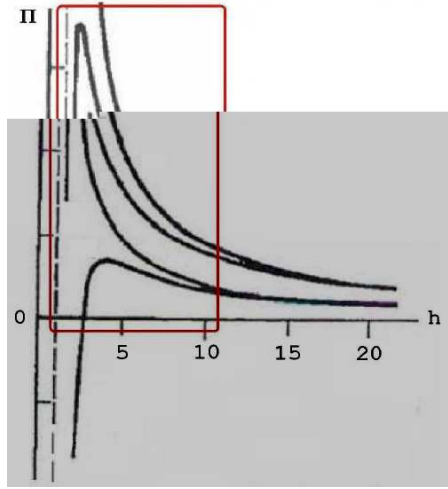}
\end{center}
\caption{\emph{Experimental curves of the disjoining pressure (issued from
Ref. \protect\cite{Derj}): the h-axis unit is the nanometer; the $\Pi $-axis
unit is about some atmospheres. The rectangle area corresponds to a domain
where the curves of the disjoining pressure (which depends on the quality of
the wall) do not have a behavior as $\,h^{-3}$ like in classical theory of
thin films \protect\cite{Lifshitz}. This experimental behavior corresponds
to nanofilms of a very few number of nanometers. This behavior is similar to
the one obtained from our model of functional which takes into account both
wall and liquid-vapor interface effects.}}
\label{fig2}
\end{figure}
Clearly, the disjoining pressure could be measured by applying an external
pressure to keep the layer in equilibrium. The measure of the disjoining
pressure is either the additional pressure on the surface or the drop in the
pressure within the \emph{mother bulks} that produce the layer. In both
cases, the forces arising during the thinning of a film of uniform thickness
$h$ produce the disjoining pressure $\Pi (h)$ of the layer with the
surrounding phases; the disjoining pressure is equal to the difference
between the pressure $P_{{v_{b}}}$ on the interfacial surface (which is the
pressure of the vapor mother bulk of density $\rho _{v_{b}}$) and the
pressure $P_{b}$ in the liquid mother bulk (density $\rho _{b}$) from which
the layer extends:
\begin{equation*}
\Pi (h)=P_{{v_{b}}}-P_{b}\,.  \label{disjoiningpressure}
\end{equation*}%
The most classical apparatus to measure the disjoining pressure is due to
Sheludko \cite{Sheludko} and is described on Fig. (1). The film is so thin
that the gravity effect is neglected across the layer. Experimental curves
of the disjoining pressure were first obtained by Derjaguin. The behavior of
the disjoining pressure for a nanofilm in \cite{Derj} seems strongly
different from the one obtained for thin liquid film in \cite{Lifshitz} (see
Fig. 2). \newline
If $g_{b}(\rho )=g_{o}(\rho )-g_{o}(\rho _{b})-\mu _{o}(\rho _{b})(\rho
-\rho _{b})$ denotes the primitive of $\mu _{b}(\rho )$, null for $\rho _{b}$%
, we get from Eq. (\ref{therm.pressure})
\begin{equation}
\Pi (\rho _{b})=-g_{b}(\rho _{v_{b}}),  \label{disjoining}
\end{equation}%
and an integration of Eq. (\ref{equilibrium2d}) yields
\begin{equation}
\frac{\lambda }{2}\,\left( \frac{d\rho }{dz}\right) ^{2}=g_{b}(\rho )+\Pi
(\rho _{b}).  \label{equilibrium2e}
\end{equation}%
The reference chemical potential linearized near $\rho _{l}$ (\emph{%
respectively} $\rho _{v}$) is $\ \mu _{o}(\rho )=\displaystyle\frac{c_{l}^{2}%
}{\rho _{l}}(\rho -\rho _{l})\ $ $\left( \emph{respectively}\ \mu _{o}(\rho
)=\displaystyle\frac{c_{v}^{2}}{\rho _{v}}(\rho -\rho _{v})\right) $ where $%
c_{l}$ (\emph{respectively} $c_{v}$) is the isothermal sound velocity in
liquid bulk $\rho _{l}$ (\emph{respectively} vapor bulk $\rho _{v}$) at
temperature $\theta $ \cite{espanet}. In the liquid and vapor parts of the
liquid-vapor film, Eq. (\ref{equilibrium2d}) yields
\begin{equation*}
\lambda \frac{d^{2}\rho }{dz^{2}}=\frac{c_{l}^{2}}{\rho _{l}}(\rho -\rho
_{b})\quad \mathrm{(liquid)}\quad \mathrm{and}\quad \lambda \frac{d^{2}\rho
}{dz^{2}}=\frac{c_{v}^{2}}{\rho _{v}}(\rho -\rho _{v_{b}})\quad \mathrm{%
(vapor)}.
\end{equation*}%
The values of $\mu _{o}(\rho )$ are equal for the mother densities $\rho
_{v_{b}}$ and $\rho _{{b}}$,
\begin{equation*}
\frac{c_{l}^{2}}{\rho _{l}}(\rho _{b}-\rho _{l})=\mu _{o}(\rho _{b})=\mu
_{o}(\rho _{v_{b}})=\frac{c_{v}^{2}}{\rho _{v}}(\rho _{v_{b}}-\rho _{v}),\ \
\mathrm{and\ consequently,}
\end{equation*}%
\begin{equation*}
\rho _{v_{b}}=\rho _{v}\left( 1+\frac{c_{l}^{2}}{c_{v}^{2}}\frac{(\rho
_{b}-\rho _{l})}{\rho _{l}}\right) .
\end{equation*}%
In the liquid and vapor parts of the complete liquid-vapor layer we get the
first expansion of the free energy, null when $\rho =\rho _{l}$ and $\rho
=\rho _{v}$ respectively,
\begin{equation*}
g_{o}(\rho )=\frac{c_{l}^{2}}{2\rho _{l}}(\rho -\rho _{l})^{2}\ \ \ \mathrm{%
(liquid)}\quad \mathrm{and}\quad g_{o}(\rho )=\frac{c_{v}^{2}}{2\rho _{v}}%
(\rho -\rho _{v})^{2}\ \ \ \mathrm{(vapor)}.
\end{equation*}%
From definition of $g_{b}(\rho )$ and Eq. (\ref{disjoining}) we deduce the
disjoining pressure
\begin{equation}
\Pi (\rho _{b})=\frac{c_{l}^{2}}{2\rho _{l}}(\rho _{l}-\rho _{b})\left[ \rho
_{l}+\rho _{b}-\rho _{v}\left( 2+\frac{c_{l}^{2}}{c_{v}^{2}}\frac{(\rho
_{b}-\rho _{l})}{\rho _{l}}\right) \right] .  \label{disjoining pressure2}
\end{equation}%
Far from the critical point, due to\ $\ \displaystyle\rho _{v}\left( 2+\frac{%
c_{l}^{2}}{c_{v}^{2}}\frac{(\rho _{b}-\rho _{l})}{\rho _{l}}\right) \ll \rho
_{l}+\rho _{b}$, we get $\displaystyle\ \Pi (\rho _{b})\approx \frac{%
c_{l}^{2}}{2\rho _{l}}(\rho _{l}^{2}-\rho _{b}^{2}).$ Now, we consider a
film of thickness $h$; the density profile in the liquid part of the
liquid-vapor film is solution of the differential equation,
\begin{equation}
\displaystyle\lambda \frac{d^{2}\rho }{dz^{2}}=\frac{c_{l}^{2}}{\rho _{l}}%
(\rho -\rho _{b})  \label{repartition}
\end{equation}
\begin{equation*}
\quad \mathrm{with}\quad \displaystyle\lambda \frac{d\rho }{dz}_{\left\vert
_{z=0}\right. }=-\gamma _{1}+\gamma _{2\ }\rho _{\left\vert _{z=0}\right.
}\quad \mathrm{and}\quad \displaystyle\lambda \frac{d\rho }{dz}_{\left\vert
_{z=h}\right. }=-\gamma _{4}\ \rho _{\left\vert _{z=h}\right. }.
\end{equation*}

With defining $\tau $ such that $\displaystyle\tau ={c_{l}}/\sqrt{\lambda
\rho _{l}}\ ,$ where $d=1/\tau $ is a reference length and $\gamma
_{3}=\lambda \tau $, the solution of Eq. (\ref{repartition}) is
\begin{equation}
\rho =\rho _{b}+\rho _{1}\,e^{-\tau z}+\rho _{2}\,e^{\tau z},  \label{profil}
\end{equation}%
where the boundary conditions at $z=0$ and $z=h$ yield the values of $\rho
_{1}$ and $\rho _{2}$ satisfying
\begin{equation*}
\left\{
\begin{array}{c}
  (\gamma _{2}+\gamma _{3})\rho _{1}+(\gamma _{2}-\gamma _{3})\rho
_{2}=\gamma _{1}-\gamma _{2}\rho _{b},  \\
  \quad\quad -e^{-h\tau }(\gamma _{3}-\gamma _{4})\rho _{1}+e^{h\tau }(\gamma
_{3}+\gamma _{4})\rho _{2}=-\gamma _{4}\rho _{b}\, .%
\end{array}%
\right. \
\end{equation*}%
The liquid density profile is a consequence of Eq. ({\ref{profil}) when $z$ $%
\in \left[ 0,h\right] $,
\begin{eqnarray}
\rho  &=&\rho _{b}+\frac{(\gamma _{1}-\gamma _{2}\rho _{b})(\gamma
_{3}+\gamma _{4})e^{h\tau }+(\gamma _{2}-\gamma _{3})\gamma _{4}\rho _{b}}{%
(\gamma _{2}+\gamma _{3})(\gamma _{3}+\gamma _{4})e^{h\tau }+(\gamma
_{3}-\gamma _{4})(\gamma _{2}-\gamma _{3})e^{-h\tau }}\ e^{-\tau z}+  \notag
\\
&&\frac{-(\gamma _{2}+\gamma _{3})\gamma _{4}\rho _{b}+(\gamma _{1}-\gamma
_{2}\rho _{b})(\gamma _{3}-\gamma _{4})e^{-h\tau }}{(\gamma _{2}+\gamma
_{3})(\gamma _{3}+\gamma _{4})e^{h\tau }+(\gamma _{3}-\gamma _{4})(\gamma
_{2}-\gamma _{3})e^{-h\tau }}\ e^{\tau z}.  \label{profil liquide}
\end{eqnarray}%
Equations (\ref{equilibrium2e},{\ref{profil}) together with $g_{b}(\rho
)=(c_{l}^{2}/2\,\rho _{l})(\rho -\rho _{b})^{2}$ for the liquid part of the
layer yield
\begin{equation}
\Pi (\rho _{b})=-\frac{2\,c_{l}^{2}}{\rho _{l}}\,\rho _{1}\,\rho _{2}.
\label{Derjaguine}
\end{equation}%
The disjoining pressure is an invariant through the liquid film and its
value is function of both $\rho _{b}$ and $h$,
\begin{eqnarray}
\Pi (\rho _{b}) &=&\frac{2c_{l}^{2}}{\rho _{l}}\left[ (\gamma _{1}-\gamma
_{2}\rho _{b})(\gamma _{3}+\gamma _{4})e^{h\tau }+(\gamma _{2}-\gamma
_{3})\gamma _{4}\rho _{b}\right] \times   \notag \\
&&\frac{\left[ (\gamma _{2}+\gamma _{3})\gamma _{4}\rho _{b}-(\gamma
_{1}-\gamma _{2}\rho _{b})(\gamma _{3}-\gamma _{4})e^{-h\tau }\right] }{%
\left[ (\gamma _{2}+\gamma _{3})(\gamma _{3}+\gamma _{4})e^{h\tau }+(\gamma
_{3}-\gamma _{4})(\gamma _{2}-\gamma _{3})e^{-h\tau }\right] ^{2}}.
\label{Derjaguine}
\end{eqnarray}%
By identification of expressions (\ref{disjoining pressure2}) and (\ref%
{Derjaguine}), we get a relation between $h$ and $\rho _{b}$. Consequently,
we get a relation between the disjoining pressure $\Pi (\rho _{b})$ and the
thickness $h$ of the liquid film. For the sake of simplicity, we denote the
disjoining pressure as a function of $h$ at temperature $\theta $ by $\Pi
=\Pi (h)$.\newline
In experiments, for liquid in equilibrium with bubbles - even with a bubble
diameter of a few number of nanometers - we have $\rho _{b}\simeq \rho _{l}$%
\ \cite{isola}. Consequently, the disjoining pressure is expressed as a
function of $h$ in the approximative form
\begin{eqnarray*}
\Pi (h) &=&\frac{2\,c_{l}^{2}}{\rho _{l}}\left[ (\gamma _{1}-\gamma _{2}\rho
_{l})(\gamma _{3}+\gamma _{4})e^{h\tau }+(\gamma _{2}-\gamma _{3})\gamma
_{4}\rho _{l}\right] \times  \\
&&\frac{\left[ (\gamma _{2}+\gamma _{3})\gamma _{4}\rho _{l}-(\gamma
_{1}-\gamma _{2}\rho _{l})(\gamma _{3}-\gamma _{4})e^{-h\tau }\right] }{%
\left[ (\gamma _{2}+\gamma _{3})(\gamma _{3}+\gamma _{4})e^{h\tau }+(\gamma
_{3}-\gamma _{4})(\gamma _{2}-\gamma _{3})e^{-h\tau }\right] ^{2}}.
\end{eqnarray*}%
\emph{Let us notice an important property} of the mixture of a fluid far
under its critical point and a perfect gas, where the total pressure is the
sum of the partial pressures of the components \cite{espanet}: at
equilibrium, the partial pressure of the perfect gas is constant through the
liquid-vapor-gas layer where the perfect gas is dissolved in the liquid. The
disjoining pressure of the mixture is the same as for a single fluid and
calculations and results are identical to those previously obtained. }}

\section{A comparison of the model with experiments}

Our aim is not to propose an exhaustive study of the disjoining pressure for
all physicochemical conditions associated with different fluids bounded by
different walls, but to point out an example such that previous modeling
appropriately fits with experimental data. At $\theta= 20 {{}^\circ}$
Celsius, we successively consider water wetting walls (a wall of silicon is
the reference material) and water not wetting a wall.

\begin{table}[tbp]
\centering
$%
\begin{tabular}{|c|c|c|c|c|c|}
\hline\hline
\multicolumn{1}{||c|}{\scriptsize Physical constants} & $c_{ll}$ & $%
\sigma_{l}$ & $m_{l}$ & $\rho_{l}$ & \multicolumn{1}{|c||}{$c_{l}$} \\ \hline
\multicolumn{1}{||c|}{Water} & $1.4\times 10^{-58}$ & $2.8\times 10^{-8}$ & $%
2.99\times 10^{-23}$ & $0.998$ & \multicolumn{1}{|c||}{$1.478\times 10^{5}$}
\\ \hline\hline
\multicolumn{1}{||c|}{\scriptsize Physical constants} & $c_{ls}$ & $%
\sigma_{s}$ & $m_{s}$ & $\rho_s$ & \multicolumn{1}{|c||}{$\delta$} \\ \hline
\multicolumn{1}{||c|}{Silicon} & $1.4\times 10^{-58}$ & $2.7\times 10^{-8}$
& $4.65\times 10^{-23}$ & $2.33$ & \multicolumn{1}{|c||}{$2.75\times 10^{-8}$%
} \\ \hline\hline\hline\hline
\multicolumn{1}{||c|}{\scriptsize Deduced constants} & $\lambda$ & $\gamma_1$
& $\gamma_2=\gamma_4$ & $\gamma_3$ & \multicolumn{1}{|c||}{$d$} \\ \hline
\multicolumn{1}{||c|}{\scriptsize Results (water-silicon)} & $1.17\times
10^{-5}$ & $81.2$ & $54.2$ & $506$ & \multicolumn{1}{|c||}{$2.31\times
10^{-8}$} \\ \hline\hline
\end{tabular}
$ \vskip 0.15cm
\caption{ The physical values associated with water and silicon are obtained
in references \protect\cite{Israel,Handbook} and expressed in \textbf{c.g.s.
units} (centimeter, gramme, second). No information is available for
water-silicon interactions; we assume that $c_{ll}=c_{ls}$. The deduced
constants are obtained from physical values by means of formulae obtained in
sections 2-4.}
\label{TableKey1}
\end{table}

Due to Eq. (\ref{profil liquide}), Fig. 3 represents water liquid density
profiles in the nanolayer. We verify the consistency of the model: \newline
$\bullet$ The density gradient is large at a few nanometer range from the
solid wall and consequently in this domain, the liquid is inhomogeneous,
\newline
$\bullet$ The boundary condition (\ref{BC2}) is well adapted to our model of
functional and determines the position where the phase transition between
liquid and vapor occurs: condition (\ref{BC2}) yields a density value of the
fluid corresponding to an intermediate density which can be associated with
a dividing surface separating liquid and vapor in the liquid-vapor
interface. Due to the film instability, we will see further down that graph
(b) in Fig. 3 is unphysical.

\begin{figure}[h]
\begin{center}
\includegraphics
[width=12cm] {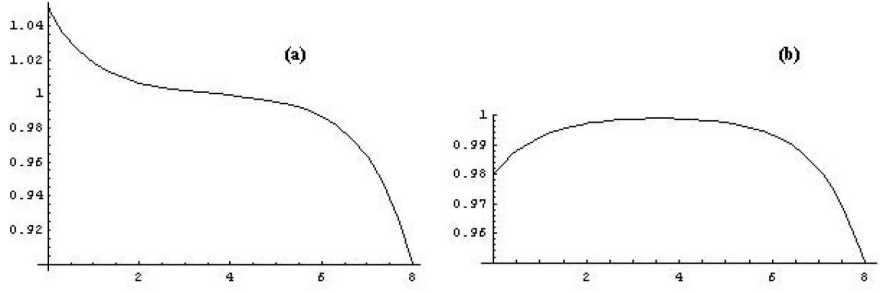}
\end{center}
\caption{\emph{Graphs of liquid density profiles in a nanofilm. Results are
given in two cases: Graph (a) corresponds to the case of a liquid damping a
solid wall ($\protect\gamma _{1}- \protect\gamma _{2}\protect\rho _{l}<0$);
wetting data are associated with liquid water on silicon (cf. graph (b) in
Fig. 4). Graph (b) corresponds to the unphysical case of non-wetting liquid (%
$\protect\gamma _{1}- \protect\gamma _{2} \protect\rho _{l} >0$);
non-wetting data are associated with liquid water and a wall such that $%
\protect\gamma _{1}=30$, all the other deduced constants $\protect\lambda,
\protect\gamma_2, \protect\gamma_3, \protect\gamma_4$ being unchanged (cf.
graph (d) in Fig. 4). The unit of the $x-$axis is $\protect\delta =
2.75\times 10^{-8}$ cm (2.75 Angstr\"{o}m), the unit of the $y-$axis is the
liquid water density at 20${{}^\circ}$Celsius (approximatively $1$ g/cm$^{3}$%
).}}
\label{fig3}
\end{figure}
We have drawn disjoining pressure profiles deduced from analytical
expressions given in section 4; the graphs relate to Rel. (\ref{Derjaguine}).%
\newline
Graphs are associated with several cases when water damps the solid wall (at
the wall, the water density is closely $\rho_l$ and $\gamma _{1}-\gamma
_{2}\rho _{l} >0$) and a case when water does not damp the solid wall ($%
\gamma _{1}-\gamma _{2}\rho _{l} <0$).

\begin{table}[tbp]
\centering
$\newline
\newline
\begin{tabular}{|c|c|c|c|c|c|}
\hline\hline
\multicolumn{1}{||c|}{Values of $\gamma_1$} & $110$ & $81.2$ & $58$ &
\multicolumn{1}{|c||}{$30$}   \\ \hline
\multicolumn{1}{||c|}{Corresponding graphs in Fig. 4} & (a) & (b) & (c) &
\multicolumn{1}{|c||}{(d)}  \\ \hline\hline
\end{tabular}
$ \vskip 0.15cm
\caption{ The numerical data $\protect\lambda, \protect\gamma_2, \protect%
\gamma_3, \protect\gamma_4$ corresponding to liquid water are unchanged.
They have the values presented in Table 1. Only the values of $\protect\gamma%
_1$ (in c.g.s. units) depending on the behavior of the solid walls are
different following graphs (a), (b), (c), (d) drawn in Fig. 4.}
\label{TableKey2}
\end{table}

\begin{figure}[h]
\begin{center}
\includegraphics[width=14cm]{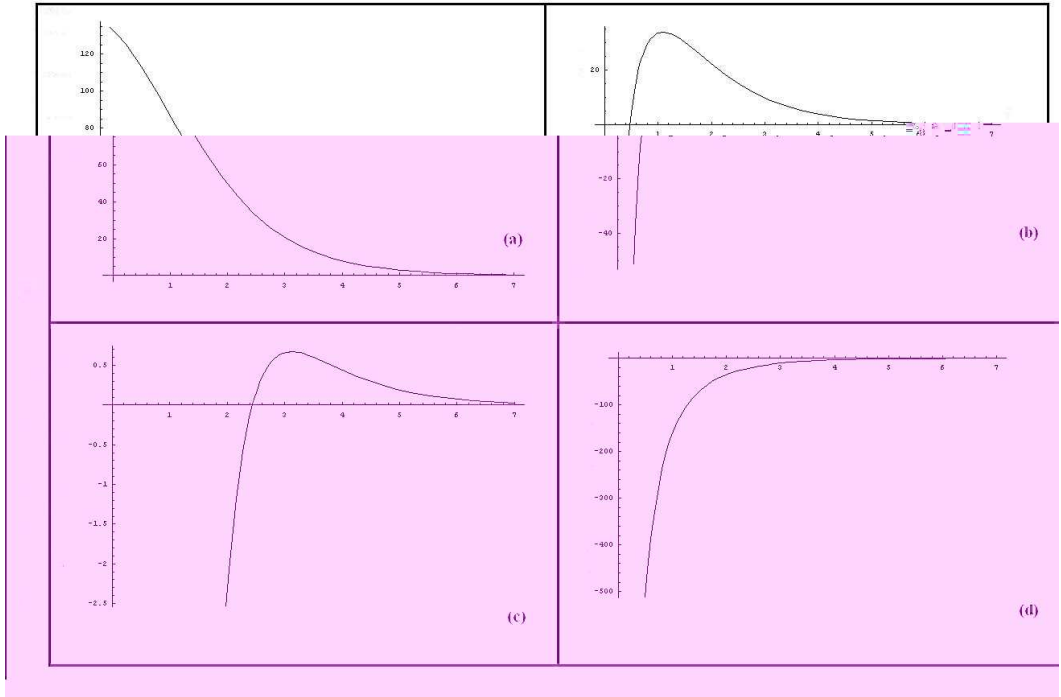}
\end{center}
\caption{\emph{Several graphs represent the disjoining pressure for liquid
water at $20{{}^{\circ }}$ C in contact with different plane solid walls:
graphs (a), (b), (c) correspond to water damping the wall; graph (d)
corresponds to an unphysical case of water not damping the wall. The unit of
the $x-$axis is $\protect\delta$, the unit of the $y-$axis is $10^6 \ Barye$
(one atmosphere).}}
\label{fig4}
\end{figure}
According to different physical values, graphs (a), (b) and (c) of Fig. (4)
represent the disjoining pressure profiles for water in contact with a plane
solid wall at $20{{}^{\circ }}$ Celsius. \newline
Graph (b) corresponds to a silicon solid wall. Graphs (a) and (c)
respectively correspond to water wetting more strongly the wall than a
silicon wall and water wetting less strongly the wall than a silicon wall.%
\newline
In \cite{gouinijes}, we studied the stability of nanofilm. In accordance
with results in \cite{Derjaguin}, Graph (a) is associated with a stable
nanolayer for any liquid film thickness because for all $\displaystyle{\ h,\
\partial \Pi (h,\theta )}/ {\partial h}<0 $. In graphs (b) and (c), values
of $h$ for which the liquid nanolayer is stable correspond to a domain where
$\displaystyle{\ \partial \Pi (h,\theta )}/{\partial h}<0 $, corresponding
to $h$ values greater than a particular value $h_s$ depending on $\gamma
_{1} $. \newline
Graph (d) differs from previous ones as that liquid water does not damp the
solid wall. The graph corresponds to an unstable nanolayer and does not
exist physically. In the non-wetting case, liquid nanolayers are unstable
and they are associated with compression instead of suction in experiments
by Sheludko \cite{Sheludko}.\newline
We notice that graphs (a), (b), (c) in Fig. 4, experimental graph in Fig. 2
and graphs in experimental literature (as in \cite{Derjaguin,Derj}) exhibit
quite similar behaviors.

\section{Conclusion}

We have studied liquid nanofilms in contact with plane solid walls. For
layer thicknesses of some nanometers, the theoretical graphs of the
disjoining pressure correctly draw the behavior of experiments by Derjaguin
and others \cite{Derjaguin,Sheludko}. The proposed analytical method is
different from Lifschitz one in which layers were considered with uniform
density liquids \cite{Lifshitz}. In our model corresponding to thin liquid
nanofilms, liquids are considered as inhomogeneous near the solid walls. The
density distribution in liquid nanofilms depends on the physicochemical
characteristics of walls: when the liquid damps the wall, we have an excess
of fluid density at the wall and the fluid is denser at the wall than in the
liquid bulk; the contrary happens when the liquid does not damp the wall.%
\newline
These analytical results and the liquid density profiles are in accordance
with experimental works by Derjaguin and others \cite%
{Derjaguin,Green-Kelly,chernov1}.

\end{document}